\documentclass[showkeys,aps,prl,twocolumn,showpacs,twoside,letterpaper]{revtex4}
\usepackage[utf8]{inputenc}
\usepackage{graphicx,amssymb,amsmath,amsbsy,subfigure,dsfont,pxfonts,pifont,ifthen,ifpdf}
\newboolean{submitToPRB} \setboolean{submitToPRB}{false}
\newboolean{draftVersion} \setboolean{draftVersion}{false}
\newcommand{\mycaption}[1]{\ifthenelse{\boolean{submitToPRB}}{\caption{(Color online) #1}}{\caption{#1}}}
\usepackage{color}
\newcommand{\ud}{\mathrm d}

\newcommand{\eqdf}{\mathop{=}\limits^{\mathrm{df}}}

\newcommand{\mybf}[1]{{\bf #1}}\renewcommand{\vec}{\mybf}

\begin{document}
\title{Coulomb-induced Rashba spin-orbit coupling in semiconductor quantum wells}
\author{Oleg Chalaev} \author{G. Vignale}
\affiliation{Department of Physics, University of Missouri-Columbia, Columbia, Missouri 65211, USA}
\date{February 17, 2010}
\begin{abstract}
In the absence of an external field, the Rashba spin-orbit interaction (SOI) in a two-dimensional electron gas in a semiconductor quantum well arises entirely from the screened electrostatic potential of ionized donors.
We  adjust the wave functions of a quantum well so that electrons occupying the first (lowest) subband conserve their spin projection along the growth axis ($s_z$), while the electrons occupying the second subband precess due to Rashba SOI.  Such a specially designed quantum well may be used as a spin relaxation trigger: electrons conserve~$s_z$ when the applied voltage (or current) is lower than a certain threshold~$V^*$; higher voltage switches on the Dyakonov-Perel spin relaxation.
\end{abstract}
\maketitle

\paragraph*{Introduction.}
The ability to control the amplitude of the Rashba\cite{RashbaSOI} spin-orbit interaction (SOI) electrically\cite{Winkler,Bastard} makes this type of SOI one of the most promising
instruments for manipulating  spins of electrons in future spintronic devices\cite{spintronics,Awschalom:2007}.
The most commonly studied case is the Rashba SOI in a two-dimensional electron gas (2DEG) inside a \emph{symmetrical} quantum well.
In such a system the Rashba SOI is zero in the absence of an external electric field, applied perpendicularly to the plane of the quantum well. Further, in a quantum well with more than one
subband, such external-field induced spin-orbit coupling does not show strong dependence on the subband index.

It is natural to ask whether Rashba SOI can also appear
in the absence of an external electric field.  A well-known argument,\cite{Winkler2004450} based on the Ehrenfest
theorem, leads to the conclusion that this cannot happen, as long as the confinement potential for the valence band is proportional to the confinement potential for the conduction band
(which is usually the case).  However, this argument fails to take into account the screened electrostatic potential  of the donors: this breaks the
proportionality between the potentials acting on electrons and holes and leads to a finite SOI in the absence of an external field.  We describe this effect as \emph{Coulomb-induced
Rashba SOI.}

An attractive feature of the Coulomb-induced Rashba SOI is that it can be made strongly subband-dependent by  a proper engineering of the shape of the quantum well.  In this letter we
use techniques of inverse scattering theory\cite{submissiveQMeng} to find a shape of the quantum well for which the first (lowest) subband is free of SOI, while the second has a rather large SOI.  We suggest
that such a shape may be realized by digital alloying techniques.  In such a specially engineered quantum well, electrons in the lowest subband would conserve their spin projection $s_z$
along the growth axis $z$ ([110]), but electrons in the second subband would suffer spin relaxation by Dyakonov-Perel mechanism.
More importantly, we suggest that a \emph{parallel} electric field (as opposed to the conventional \emph{perpendicular} electric field)
might be used to control the SOI amplitude by ``pumping'' electrons into the second subband.
The electrostatic potential generated by this non-equilibrium population would ``turn-on'' the Rashba coupling, leading to Dyakonov-Perel spin relaxation also in the lowest subband.

\begin{figure}
\includegraphics[width=.6\columnwidth]{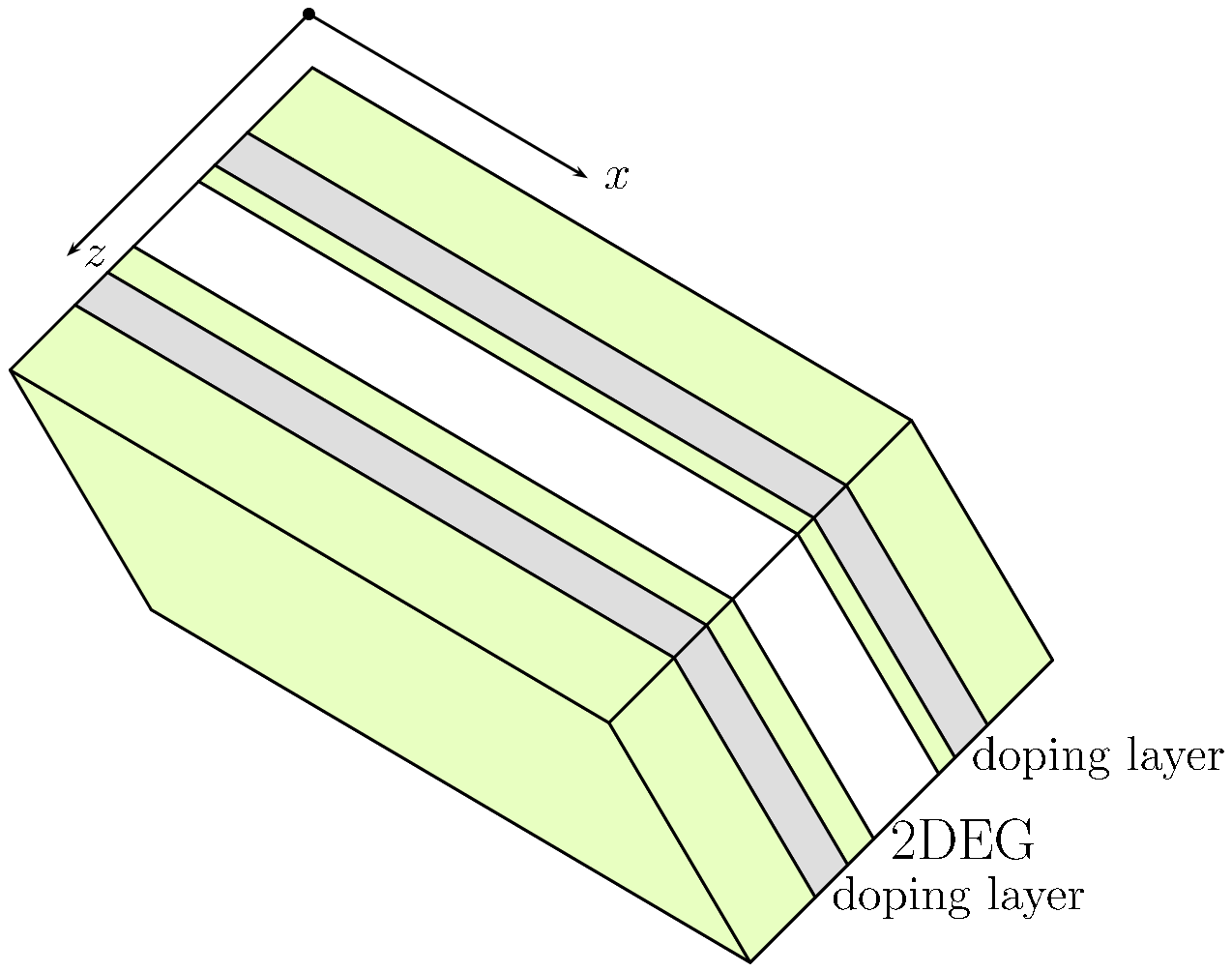}
\mycaption{The proposed device. The 2DEG is confined inside the quantum well grown along the $z$- ($[110]$) direction; the current flows along the $x$-direction.
The widths of the spacers separating the central zone (the well) from the doping layers on the two sides of the well are different.
\label{fig:device}}
\end{figure}

\paragraph{The model.}
We consider a two-dimensional electron gas confined within a quantum well (QW) in an Al$_x$Ga$_{1-x}$As heterostructure (see Fig.~\ref{fig:device}).

We assume that the electrons in the QW are provided by donors outside the well, and that the concentration of donors is high enough
so that the 2DEG is degenerate.  The electronic states in the well are plane waves with momentum $\vec p$ in the plane of the well $(x,y)$.  They are further characterized by a subband
index $n$, which determines the transverse envelope wave function $\varphi_n(z)$.  Our QW contains only two subbands for electrons, $n=1$ or $2$, and only the lowest subband ($n=1$) is
populated in equilibrium.

If the quantum well is grown along the [110] direction, the Dresselhaus\cite{Dresselhaus} spin-orbit interaction is absent, and the main source of spin relaxation is the Rashba\cite{spinRelax} spin-orbit interaction,
\begin{equation}
H_{\mathrm R} = \alpha_n(\sigma_x\hat p_y-\sigma_y\hat p_x),
\end{equation}
where the  SOI amplitude $\alpha_n$ is given by
\begin{equation}\label{da2}
 \alpha_n=\frac{P^2}{3\hbar}\left\langle n\left|\partial_z{\mathbb U}_{\mathrm v}\right|n\right\rangle\left[\frac1{E_g^2}-\frac1{(E_g+\Delta)^2}\right],
\end{equation}
${\mathbb U}_{\mathrm v}$ is the \emph{complete}\cite{EguesLoss} potential acting on the holes\cite{Winkler}, $E_g=1.52\,\mathrm{eV}$ is the fundamental band gap,
$\Delta=0.34\,\mathrm{eV}$, is the spin-orbit splitting of the valence band, $P=10.5\,\mathrm{eV}\!\!\cdot\!\!\text{\AA}$ is the matrix element of momentum between $s$ and $p$
atomic states, and $|n\rangle$ are subband envelope functions.  Apart from the bare valence band confinement $E_{\mathrm v}$ and a possible external potential $U_{ext}$ (such as the one
that arises from a perpendicular electric field), ${\mathbb U}_{\mathrm v}$ includes the self-consistent ``Hartree'' potential $U_{\mathrm H}$ induced by the donors and the inhomogeneous
charge distribution inside the well:
\begin{equation}\label{HartreePot}
 {\mathbb U}_{\mathrm v}(z)=U_{\mathrm{ext}}+E_{\mathrm v}(z)+U_{\mathrm H}(z).
\end{equation}
For $x<0.45$, the confinement potential in the valence band of Al$_x$Ga$_{1-x}$As is proportional\cite{Winkler,Davies} to the confinement potential, $E_c(z)$, in the conduction band:
\begin{equation}\label{EvPropToEc}
 E_{\mathrm v}(z)=2E_{\mathrm c}(z)/3.
\end{equation}

Setting $U_{ext}=0$ and making use of Eq.~(\ref{EvPropToEc}) we can write 
\begin{equation}
{\mathbb U}_{\mathrm v}=\frac23{\mathbb U}_{\mathrm c}(z)+\frac53U_{\mathrm H}(z),\quad {\mathbb U}_{\mathrm c}(z)=E_{\mathrm c}(z)-U_{\mathrm H}(z),
\end{equation}
where ${\mathbb U}_{\mathrm c}(z)$ is the full potential acting on the electrons. 
The expectation value of the force $\partial_z{\mathbb U}_{\mathrm c}$ in the state $|n\rangle$ vanishes by the Ehrenfest theorem,
so that only the contribution from the Hartree potential survives.  
We conclude that, in the absence of external fields, the Rashba SOI is produced \emph{entirely} by the Coulomb contribution~$U_{\mathrm H}$ to the hole potential,
\begin{equation}\label{finalSOIamp}
    \alpha_n=B\left\langle n\left|\partial_zU_{\mathrm H}\right|n\right\rangle,\quad B=\frac{5P^2}{9\hbar}\left[\frac1{E_g^2}-\frac1{(E_g+\Delta)^2}\right],
\end{equation}
where~$U_{\mathrm H}$ is obtained from the self-consistent solution of both Schr\"odinger and  Poisson equations for the electron envelope functions~$\varphi_n(z)$.
It is evident from~\eqref{finalSOIamp}  that the SOI vanishes in all subbands if the confinement potential $E_{\mathrm c}(z)$ and the donor distribution $\rho_D(z)$ are symmetric.  The reason is that in this case $\partial_z U_{\mathrm H}(z)$ is an odd function of $z$, while the envelope function $\varphi_n(z)$ has a definite parity.  

\paragraph{Spin-orbit coupling engineering.}
When the confinement potential~$E_{\mathrm c}$ and/or the donor distribution~$\rho_D$ is asymmetric, both SOI  amplitudes $\alpha_1$ and $\alpha_2$ are, in general, non-zero.
In order to eliminate the Dyakonov-Perel relaxation in equilibrium, we have to engineer a confinement~$E_{\mathrm c}$ for which\cite{endnote11} $\alpha_1=0$ but $\alpha_2\ne0$.
This can not be achieved by breaking the symmetry of either~$E_{\mathrm c}$ or~$\rho_D$; one has to break \emph{both} symmetries, so that
\begin{equation}\label{breakeBoth}
  E_{\mathrm c}(-z)\ne E_{\mathrm c}(z)\quad\text{and}\quad\rho_D(-z)\ne\rho_D(z).
\end{equation}
We proceed as follows: first we introduce a slight asymmetry in the positioning of the doping layers with respect to the center of the well, thus breaking the symmetry of~$\rho_D(z)$ (see
Figs.~\ref{fig:device} and~\ref{fig:nonSym}).
On top of that, we utilize a technique of inverse scattering theory known as {\it Double Darboux Transformation} (DDT),\cite{submissiveQMeng}
to generate an asymmetric confinement potential $E_c(z)$, which satisfies the first of 
inequalities~\eqref{breakeBoth}.

In a DDT one changes the single particle potential $E_{\mathrm c}^{(0)}\to E_{\mathrm c}$
and the wave functions $\varphi_n^{(0)}\to\varphi_n$ in such a way that the eigenvalues $\epsilon_n$ are preserved.
  This, in turn, offers a way to significantly modify the strengths of the SOI amplitudes without grossly altering
the spectrum of the quantum well.  Naturally, the invariance of the eigenvalues under the DDT is rigorous only if electron-electron interactions are neglected. However, we have found that, even in the presence of interactions, the variation of the eigenvalues remains relatively small, about $10\%$.

The precise form of the transformation is\cite{submissiveQMeng}
\begin{equation}\label{specPa}\hspace{-3ex}
  \begin{split}
    E_{\mathrm c}=E_{\mathrm c}^{(0)}-2\left[\varphi_2A\right]',\ \varphi_1=\varphi_1^{(0)}-AI_{21},\ \varphi_2=\frac{rA}{r^2-1}, &\\
  A=\frac{\left(r^2-1\right)\varphi_2^{(0)}}{1+\left(r^2-1\right)I_{22}},\quad I_{nm}(z)=\int_{-\infty}^z\varphi_n^{(0)}(y)\varphi_m^{(0)}(y)\ud y,&
  \end{split}\end{equation}
where $E_{\mathrm c}^{(0)}$ is the original (symmetric) confinement, and $\varphi_n^{(0)}$ are the corresponding wave functions (WFs).
The parameter $R\equiv\log r$ measures the strength of the symmetry breaking of $E_{\mathrm c}^{(0)}$;
$R=0$ corresponds to the identity transformation, i.e., $E_{\mathrm c} = E_{\mathrm c}^{(0)}$ and $\varphi_n = \varphi_n^{(0)}$.
We assumed in~\eqref{specPa} that $\varphi_{1,2}(-\infty)=+0$ and $\int_{-\infty}^\infty \varphi_n^2(z)\ud z=1$.
We further assume that $E_{\mathrm c}^{(0)}(z)$ represents a rectangular well parameterized by $U_0$ (height) and $a$ (width).

\begin{figure}
\ifthenelse{\boolean{draftVersion}}{\includegraphics[width=\textwidth]{2_levels/figures/SOI1}}{\includegraphics[width=.8\columnwidth]{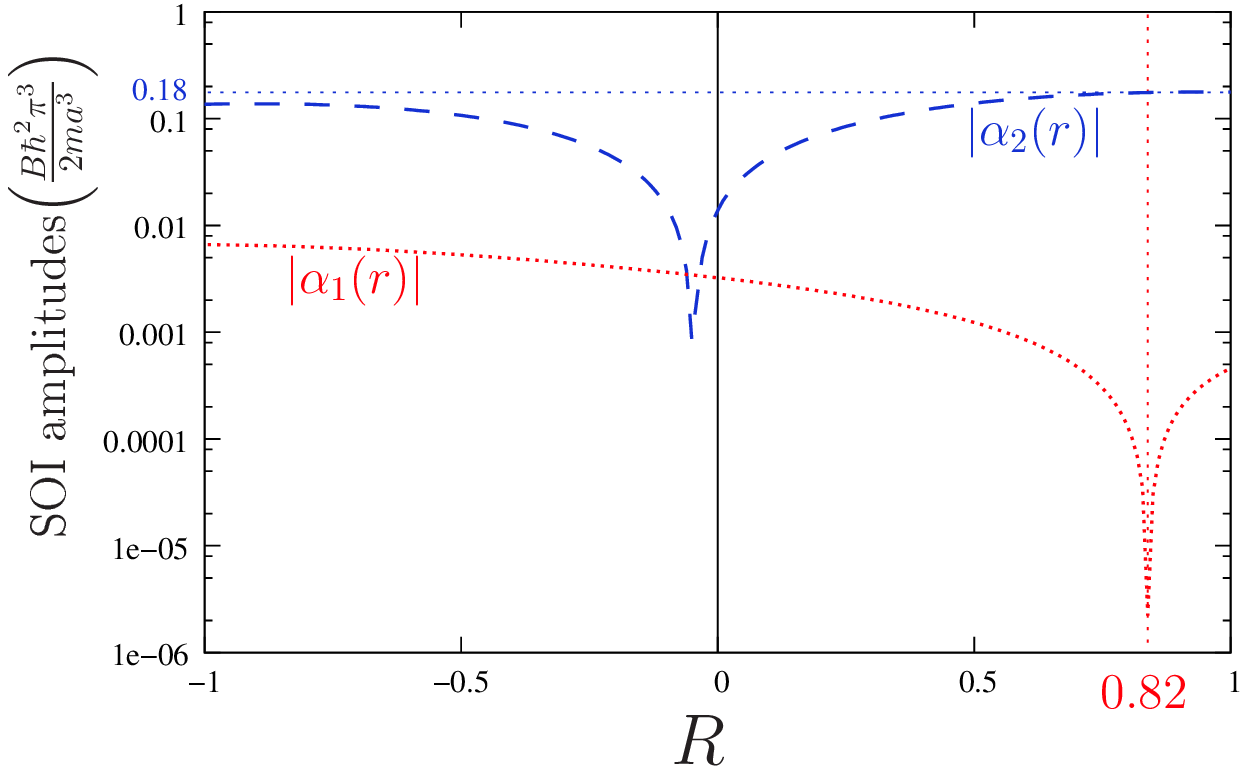}}
\mycaption{The dependence of the SOI amplitudes for electrons  in the first and second subband on the DDT-parameter $R$, see~\eqref{specPa}.   The initial confinement potential, at $R=0$,
  is a square well, with parameters described in the caption of Fig.~\ref{fig:nonSym}.
Observe that $\alpha_1=0$ at $R=0.82$. The curves are numerical\cite{www} so that zeros are not exactly reached.\label{fig:SOI}
}
\end{figure}

In Fig.~\ref{fig:SOI} we show the dependence of $\alpha_1$ and $\alpha_2$ on the parameter $R$.  We see that $\alpha_1 = 0$
when $R=0.82$.  At this value of $R$ we have $\alpha_2=0.18$ in the chosen units.  The form of the confinement potential at this value of $R$ is shown in
Fig.~\ref{fig:nonSym}, together with the plot of the electron envelope functions for the two subbands.
Such a potential might  be realized by MBE techniques.\cite{MBE}
When only the lowest subband in this specially engineered QW is populated, the electrons  do not suffer spin relaxation.
This happens in equilibrium, when the energy distribution of electrons everywhere in the sample is given by a Fermi function.
When the sample is driven out of equilibrium (e.g., by the current or by light illumination), so that both bands are populated, the SOI effects are activated.
\begin{figure}
\ifthenelse{\boolean{draftVersion}}{\includegraphics[width=.8\columnwidth]{2_levels/figures/outside}}{\includegraphics[width=.8\columnwidth]{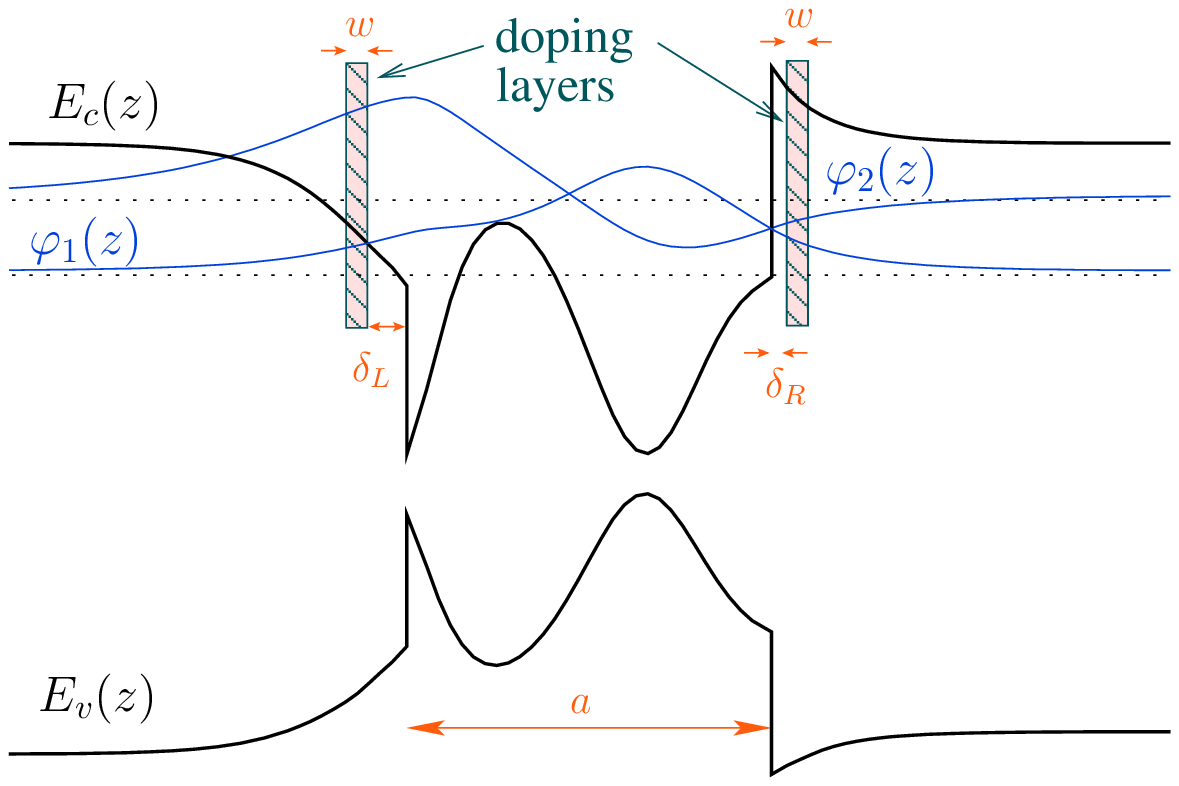}}
\mycaption{Confinement in the conduction and valence bands at $R=0.82$, see  Fig.~\ref{fig:SOI}.
The vertical walls correspond to the borders of the initial square well of width $a=400\text{\AA}$ and depth $U_0=10\mathrm{meV}$.  The left and right spacers have widths $\delta_L=0.1a$
and  $\delta_R=0.04a$, respectively. Both doping layers have widths $w=0.06a$.
The  two lowest subbands are at energies $\epsilon_1=4.5\mathrm{meV}$ and $\epsilon_2=8.8\mathrm{meV}$ from the bottom of the original well.
\cite{www}\label{fig:nonSym}}
\end{figure}

\paragraph{Changing the population of the subbands.}
It is well-known\cite{Rashba} that the Rashba SOI amplitude can be controlled by a \emph{perpendicular} (to the QW plane) electric field.  Below we  show that an
electric field which drives a current in the plane of the QW, can be used as a ``switch'' for the SOI amplitude, when this amplitude is initially set to zero by the method described in the previous section.

Under ordinary conditions, when a current flows in an electron gas, in the linear response regime, strong inelastic interactions among the electrons and between electrons and phonons establish a local equilibrium
distribution, controlled by a local (position-dependent) electrochemical potential $\mu(\tilde x)$:
\begin{equation}\label{muOfx}
\mu(\tilde x)=(1-\tilde x)\mu(0)+\tilde x\mu(1),\quad\tilde x\eqdf x/L\,,
\end{equation}
where $x$ is the direction along the current, $L$ is the sample length,
$\mu(0)\equiv \mu_L=\mu$ and $\mu(1)\equiv \mu_R=\mu(0)+eV$ are the electrochemical potentials in the left and right contacts, and $V$ is the applied voltage.  Similar to $\mu(\tilde x)$, also the subband energies  and the density of states depend  on the coordinate,
\begin{equation}\begin{split}
    \epsilon_{1,2}(\tilde x)&=\epsilon_{1,2}(0) +\mu(\tilde x)-\mu(0),\quad\epsilon_{1,2}(0)\equiv\epsilon_{1,2},\\
\nu(\tilde x,E)&=
\begin{cases}
  0,\quad E<\epsilon_1(\tilde x),\\
\nu_0,\quad \epsilon_1(\tilde x)<E<\epsilon_2(\tilde x),\\
2\nu_0,\quad  E>\epsilon_2(\tilde x)\,,\\
\end{cases}  \end{split}\end{equation}
where $\nu_0=m/(2\pi\hbar)$, so that the concentration of electrons $n$ remains independent of position.

\begin{figure}
\ifthenelse{\boolean{draftVersion}}{\includegraphics[width=.8\columnwidth]{2_levels/figures/explain}}{\includegraphics[width=.95\columnwidth]{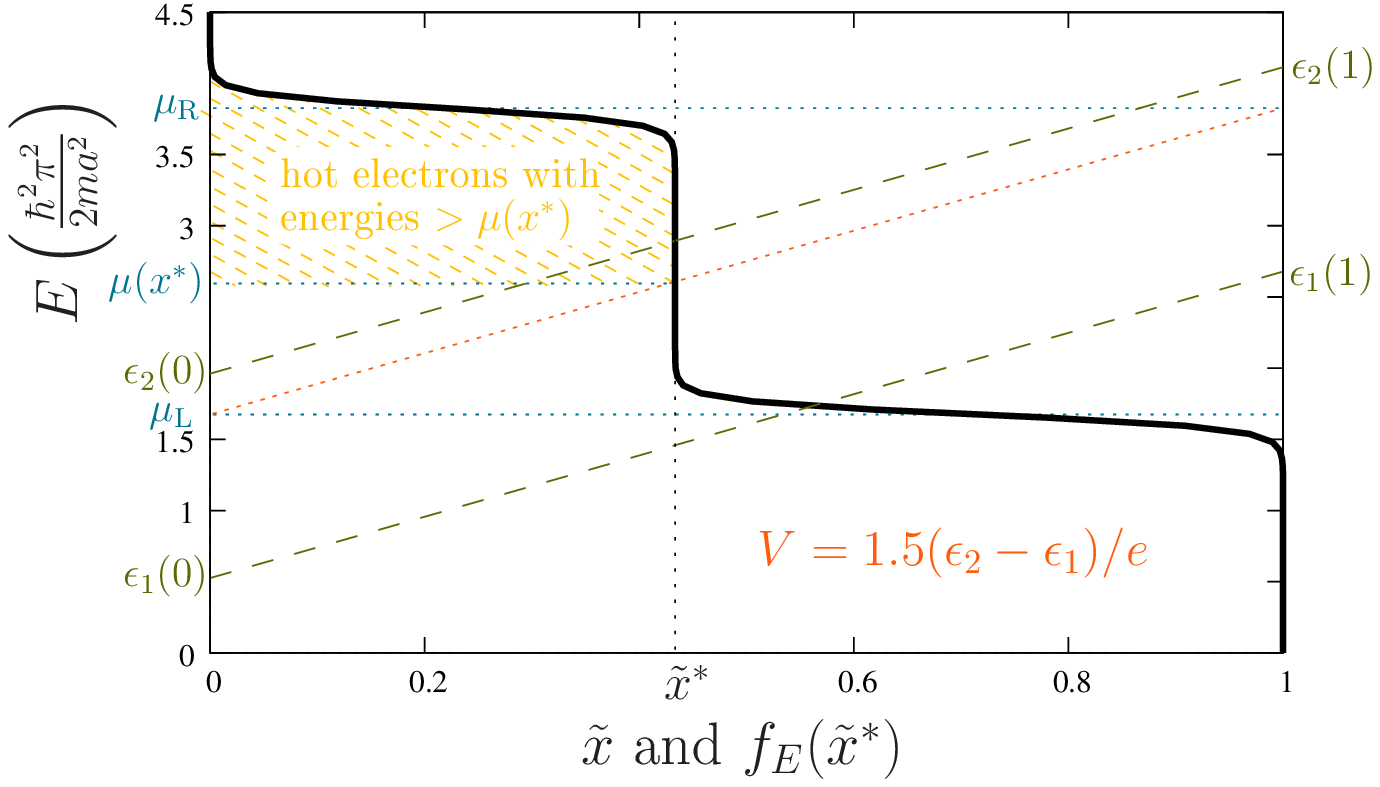}}
\mycaption{The energy distribution function, $f_E({\tilde x}^*)$ at a point ${\tilde x}^*$ inside a current-carrying sample (see Eq.~\eqref{lcED}) at temperature
$T=(\epsilon_2-\epsilon_1)/30$.  The dashed lines represent the edges of the conduction subbands, and the dotted line is the local electrochemical potential from Eq.~(\ref{muOfx}).
Because of the two-step form of the non-equilibrium distribution function some electrons have energy higher than the chemical potential $\mu({\tilde x}^*)$ (see the
``dashed'' area).  These ``hot'' electrons came from the right; the ``hottest'' ones have energy $\sim\mu_R$ and originate from the right contact.
Note that $f_E({\tilde x}^*) \simeq 1$ for $E \lesssim \mu_L$ and $f_E({\tilde x}^*) \simeq \tilde x^*$ for $\mu_L \lesssim E \lesssim \mu_R$: thus, the step in the distribution function
$f_E({\tilde x}^*)$ occurs precisely at $f_E({\tilde x}^*)={\tilde x}^*$.  By applying this rule one easily sees that the energy distributions at the edges of the sample, $f_E(0)$ and
$f_E(1)$, are usual Fermi functions.\label{fHotElectron}}
\end{figure}

\begin{figure}
\ifthenelse{\boolean{draftVersion}}{\includegraphics[width=.8\columnwidth]{2_levels/figures/population}}{\includegraphics[width=.8\columnwidth]{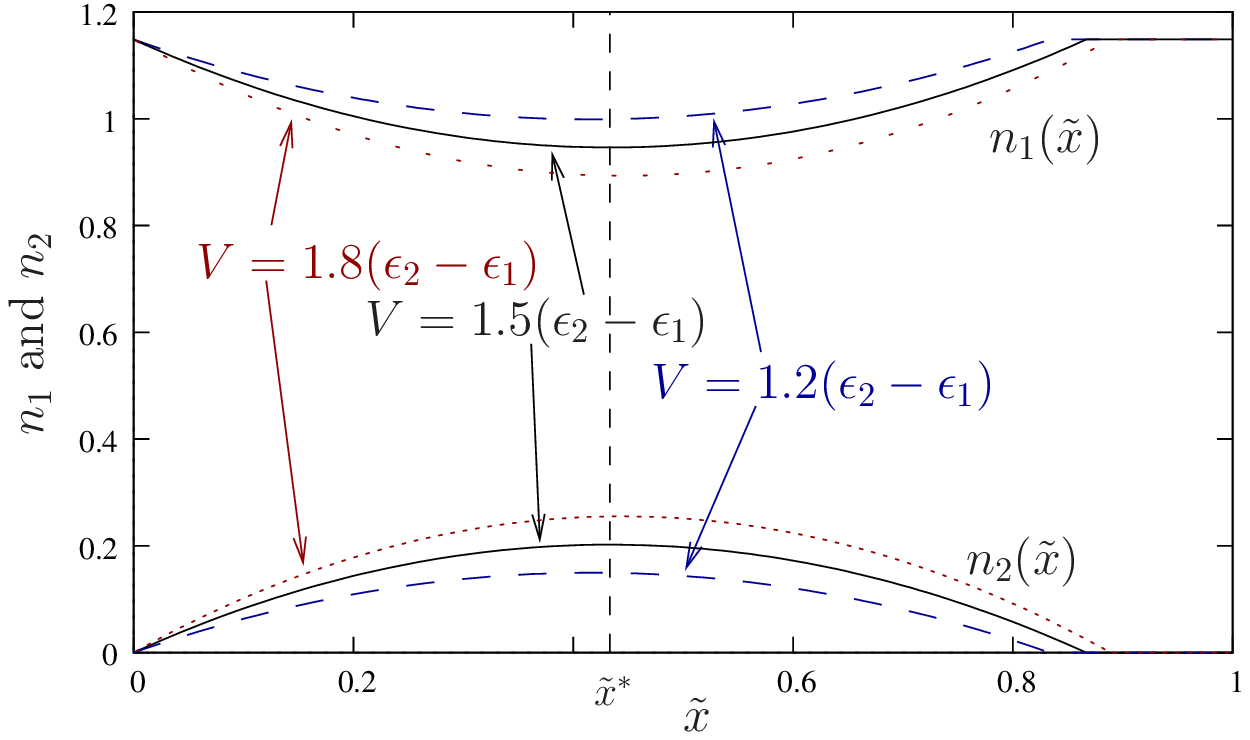}}
\mycaption{Population of the subbands in the sample, see Eq.~\eqref{population}. The width of the populated zone is controlled by the applied voltage~$V$.
For $V=1.5(\epsilon_2-\epsilon_1)$, $n_2(x)$ approaches its maximal value at ${\tilde x}^*\equiv x^*/L=0.43$.\cite{www}\label{fig:population}}
\end{figure}

However, in a mesoscopic sample at small temperatures and high mobility values, when the condition $D/L^2\gg\tau_{in}^{-1}$ is satisfied, where $D$ is the diffusion constant and $\tau_{in}$
is the inelastic scattering from electron-electron and electron-phonon interactions, the energy relaxation effects are so small that an electron may pass through the entire sample
conserving its energy.
This results in a two-step form of an energy distribution function inside the sample,\cite{Pothier,Kamenev2} see Fig.~\ref{fHotElectron}:
\begin{equation}\label{lcED}
f_E(\tilde x)=\left(1-\tilde x\right)f_E(0)+\tilde xf_E(1),
\end{equation}
where $f_E(0)$ and $f_E(1)$ are Fermi energy distributions in the left and right contacts attached to the sample, respectively.
When the voltage $V$ across the sample exceeds the critical value $V^*=(\epsilon_2-\epsilon_1)/e$, the energy of the ``hot'' electrons from the right contact exceeds the energy of the bottom of the
second subband on the left  end of the device.
The fraction of such ``hot'' electrons is controlled by~\eqref{lcED}: it is maximal near the right contact and vanishes in the vicinity of the left contact.
More precisely,  the $\tilde x$-dependent population of the second subband is (see Fig.~\ref{fig:population})
  \begin{equation}\label{population}
    \frac{n_2(x)}{\nu_0}=
\begin{cases}
    0, & \mathrm{when}\quad \epsilon_2(\tilde x)>\mu(1), \\
    \tilde x\left[\mu(1)-\epsilon_2(\tilde x)\right]/2, & \mathrm{when}\quad \epsilon_2(\tilde x)\le\mu(1)\,,
\end{cases}
  \end{equation}
where $n_1(\tilde x)=n-n_2(\tilde x)$, and $n$ is  the total density of electrons.
In~\eqref{population} we assumed that electrons spread equally between the bands when their energy allows them to be either in first or the second subband.

\begin{figure}
\ifthenelse{\boolean{draftVersion}}{\includegraphics[width=.8\columnwidth]{2_levels/figures/SOI_vs_coordinate}}{\includegraphics[width=.8\columnwidth]{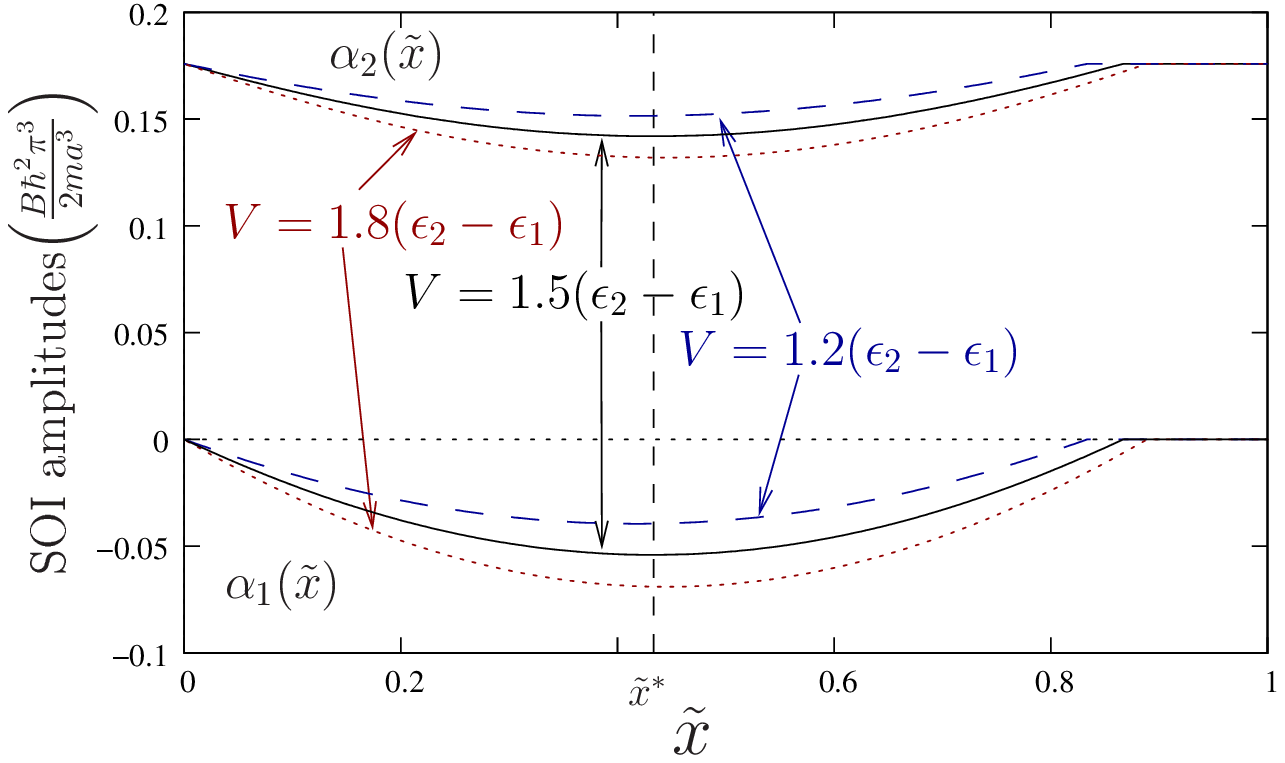}}
\mycaption{The coordinate dependence of the SOI amplitudes inside the sample. 
Close to the right contact, the second subband is not populated (see Fig.~\ref{fig:population}); hence $\alpha_1=0$, and  electrons do not
  experience spin-orbit scattering there.\cite{www}\label{fig:SOIvsCoordinate}}
\end{figure}

In equilibrium, the second subband would remain unpopulated everywhere in the sample since $\mu(\tilde x)<\epsilon_2(\tilde x)$ for arbitrary $\tilde x$.
In a steady (non-equilibrium) current-carrying state, there is a zone inside the sample, where \emph{both} subbands are populated. The width of this zone is controlled by the
applied voltage, see Fig.~\ref{fig:population}.
The electric field-induced change in the population of the subbands modifies the Hartree potential~$U_{\mathrm H}$, which in turn modifies the values of $\alpha_1$ and $\alpha_2$.  In
particular $\alpha_1$ becomes non-zero when the voltage $V$ exceeds a critical value $V^*$ ($V^*=0.8\mathrm{mV}$ for 
$\mu(0)-\epsilon_1=3.5\mathrm{meV}$), with the other parameters being specified in the caption of Fig.~\ref{fig:nonSym}.
The coordinate dependence of $\alpha_1$ and $\alpha_2$ for different voltages $V$ is shown
in Fig.~(\ref{fig:SOIvsCoordinate}).
We conclude that the Dyakonov-Perel spin relaxation of electrons in the lowest (majority) subband can be switched on by an electric field (or a current) in the plane of the QW.  It is
essential for the argument that this field does not modify the envelope functions of the electrons.  The field affects $\alpha_n$  due to the subband repopulation, and the
effect is large only because $\alpha_1$ had previously been fine-tuned to be zero.

\paragraph{Discussion.}
The method described in this Letter allows in principle to quickly suppress a spin polarization by applying a current.  A crucial point is the feasibility of keeping the hot electrons
hot (i.e., to avoid thermalization).   One way to ensure this is to use short wires, low temperatures and high mobilities, so that the condition $D/L^2\gg\tau_{in}^{-1}$ is satisfied.
Failing this, another possibility would be to rely on the nonlinear hot electron effect, see p.120 in~\cite{BlBuPR00}, whereby, in the presence of strong inelastic scattering, the distribution at the center of the sample is a Fermi distribution with an effective temperature $T^*=\left[T^2+\frac{3}{4\pi^2}(eV)^2\right]^{1/2}$.  If $V$ is sufficiently large, this will produce the desired population of the higher subband.
Finally, we notice that, while our treatment has neglected the intrinsic Dresselhaus SOI,  the latter's action may be compensated by a Rashba SOI having the same amplitude,
so that ${\hat s}_x-{\hat s}_y$ becomes a conserved quantity.\cite{RashaEqDresselhaus} So the proposed scheme should work even in the presence of Dresselhaus spin-orbit interaction.

We acknowledge the support of ARO Grant No. W911NF-08-1-0317.
We thank Michael Flatté for helpful discussions.

\bibliography{refs.aps,books.aps,local}

\begin{thebibliography}{20}
\bibitem[{end()}]{endnote11}
\bibinfo{note}{In a situation when $a\ll w$ and $a\ll\delta_{R,L}$, there is an
  important contribution to the SOI amplitude due to microscopic fluctuations
  of the donor concentration\cite{PhysRevB.71.241312}. In our case this
  contribution is estimated to be extremely small, $\sim2\times10^{-6}$
  relative to the usual Rashba contribution.}
\end{thebibliography}
\end{document}